\documentclass[aps,prl,twocolumn,superscriptaddress,10pt]{revtex4-1} 
\usepackage{braket}
\usepackage{graphicx} 
\usepackage{dcolumn}
\usepackage{bm}
\usepackage{amsmath,latexsym,tabularx}
\usepackage{multirow}
\usepackage[export]{adjustbox}
\usepackage{hyperref}
\usepackage{upgreek}
\hypersetup{
  colorlinks=true,
  urlcolor=blue,
  linkcolor=blue,
  citecolor=blue
}

\usepackage[inkscapelatex=false]{svg}

\newcommand{\affilUMass}[0]{Department of Electrical and Computer Engineering, University of Massachusetts Amherst, Amherst, MA 01003}

\newcommand{\affilUCSB}[0]{Department of Electrical and Computer Engineering, University of California Santa Barbara, Santa Barbara, CA 93106}

\newcommand{\Sr}{$^{88}$Sr$^+${~}}
\newcommand{\SiN}{$\text{Si}_3 \text{N}_4~$}

\begin{document}

\title{Trapped Ion Qubit Gates and Entanglement Driven by an Integrated Photonic Laser}

\author{Christopher Caron$^1$$^*$, Zhenyu Wei$^1$$^*$, Andrei Isichenko$^2{}$, David Heim$^2{}$, Meiting Song$^2{}$, Nitesh Chauhan$^2{}$, Nick Montifiore$^2{}$, Kaikai Liu$^2$,  Josiah Dill$^1$, Daniel J. Blumenthal$^2{}^{\dag}$, Robert J. Niffenegger$^1{}^{\dag}$
\\ \normalsize $^{1}$\affilUMass
\\ \normalsize $^{2}$\affilUCSB
\\ \normalsize $^{\dag}$Corresponding Authors: rniffenegger@umass.edu, danb@ucsb.edu
\\ \normalsize $^*$These authors contributed equally \\ 
}

\date{\today}

\begin{abstract}
Trapped ions are a leading technology for quantum computing, but their reliance on bespoke tabletop laser and optical systems remains a major obstacle to scaling and robustness. Integrated silicon nitride lasers, compatible with future monolithic integration with surface electrode ion traps, have recently demonstrated frequency-selective qubit state preparation and measurement as well as interrogation of an optical clock transition. 
However, coherent qubit gates impose even more stringent requirements on laser phase noise, and their implementation with integrated laser sources has remained an outstanding challenge. Here, we use a visible-wavelength integrated Brillouin laser stabilized to an integrated coil resonator to drive coherent single- and two-qubit gates with \Sr optical qubits. We measure an average single-qubit fidelity of $99.61\%\pm 0.03\%$ per Clifford gate with randomized benchmarking and use a two-qubit Mølmer-Sørensen interaction to generate an entangled Bell state with a fidelity of $92.35\% \pm 1.50\%$. 
The qubit exhibits a bare Ramsey coherence time of $659 \pm 9~\mu$s, more than a tenfold improvement over our previous implementation, which extends to $1.750 \pm 0.033$~ms with an echo. 
These results demonstrate, for the first time, that integrated visible-wavelength narrow-linewidth photonic lasers can meet the phase-noise requirements for coherent quantum logic with trapped ion optical qubits, providing a path for scalable optical systems integrated within trapped ion quantum processors.

\end{abstract}

\maketitle

\section{Introduction}
\vspace{-0.2in}

Trapped-ion qubits have demonstrated high fidelity single-qubit gates\cite{smith2025single}, two-qubit gates\cite{hughes2025trapped}, and state preparation and measurement\cite{sotirova2024high} as well as large quantum volumes\cite{ransford2025helios}.
However, these systems rely on complex tabletop laser systems, bulk optics, and environmentally sensitive ultra-low-expansion (ULE) optical reference cavities, which are barriers to scaling trapped ion quantum computers\cite{bruzewicz2019trapped}. Developing compact, low-noise, narrow-linewidth, stabilized laser systems and optical qubit control is a central challenge in scaling trapped-ion quantum technologies. 
Future photonic integration of the laser systems with the ion trap has the potential to substantially reduce relative phase and polarization noise between the laser source and ion qubits by replacing the long fiber and free-space optical paths with fixed on-chip waveguide routing and grating coupler delivery\cite{niffenegger2020integrated, mehta2016integrated,mehta2020integrated}. 
This would alleviate an important challenge associated with phase sensitive optical qubits\cite{bruzewicz2019trapped} and is a key motivation for using materials like \SiN which are compatible with monolithic integration into the ion trap chip without second harmonic generation.
Integration could also increase the optical intensity delivered to the trapped ions enabling faster qubit gates and potentially improving gate fidelity. 

Integrated photonic technologies have made progress toward integrating the optical systems required for trapped-ion qubits, including optical beam delivery\cite{mehta2016integrated, mehta2020integrated, niffenegger2020integrated, kwon2024multi, mordini2025multizone,morrison2025autonomous}, modulation\cite{hogle2023high}, and narrow-linewidth laser stabilization\cite{chauhan2026chip,loh2025optical}.
In terms of laser stabilization, visible\cite{chauhan2026chip} and infrared\cite{loh2025optical} integrated coil resonators have been used as frequency references enabling interrogation of narrow optical clock transitions in trapped ions without requiring bulk optic ULE reference cavities. Visible light coil resonators have also demonstrated direct, high fidelity qubit state preparation and measurement (SPAM)\cite{chauhan2026chip}. 
However, coherent single-qubit and two-qubit gates driven by integrated laser sources have not yet been demonstrated, as coherent quantum logic imposes substantially more stringent requirements than spectroscopy or qubit state preparation and measurement. Single-qubit gates require optical phase coherence throughout the pulse sequence and motion-mediated two qubit gates require low phase noise near the trapped ion motional frequencies. 

\begin{figure*}
    \centering
    \includegraphics[width=\linewidth]{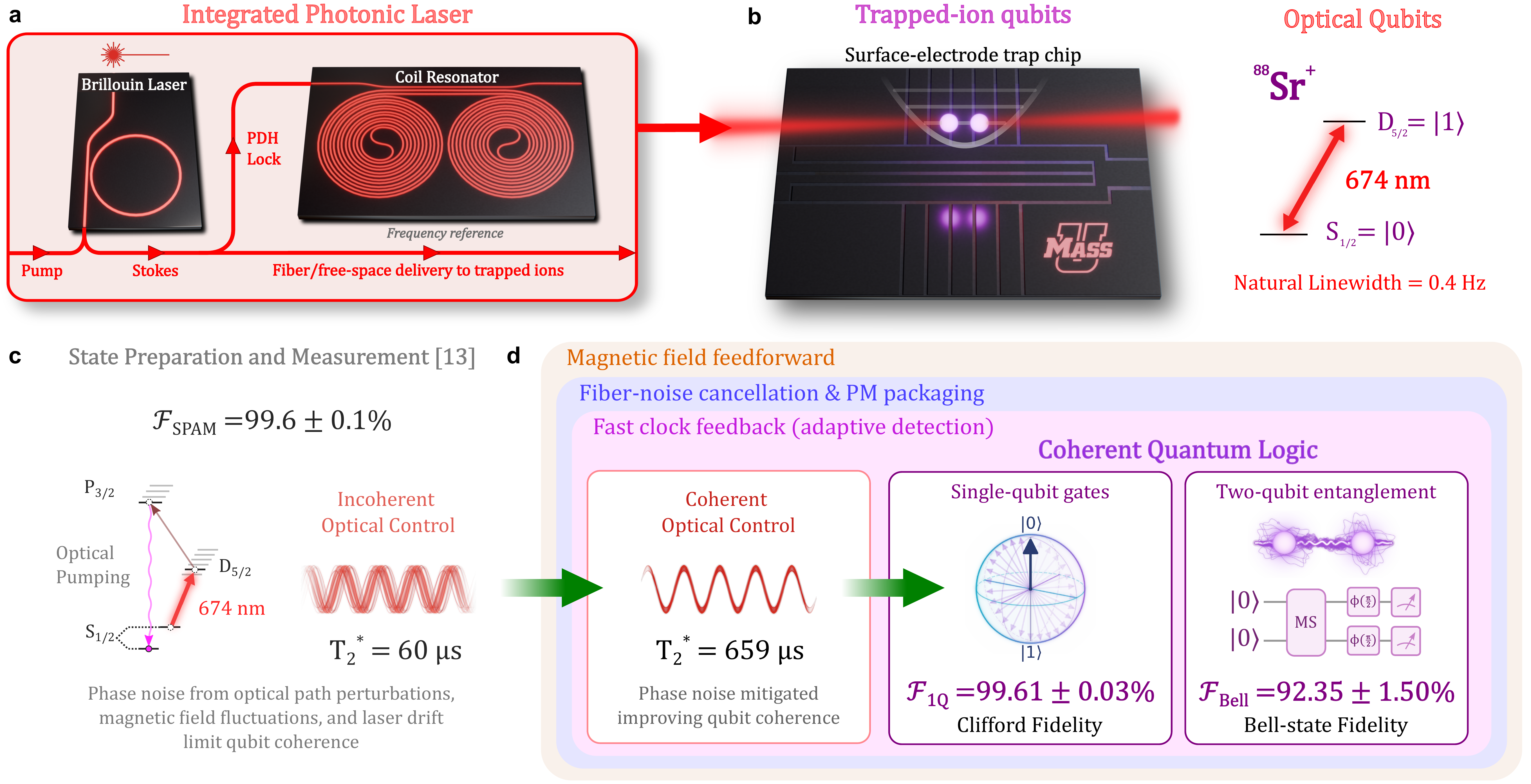}
	\caption{
    \textbf{Integrated photonic laser control of trapped ion qubits.} 
    \textbf{a}  Integrated photonic laser comprised of a Brillouin laser locked to an integrated coil resonator and delivered to the trapped ion qubits.
    \textbf{b} Two ions trapped above a surface electrode trap and coherently addressed by the integrated 674 nm laser. Energy levels of \Sr trapped ion optical qubit encoded in the $5S_{1/2}$ ground state ($\ket{0}$), and metastable $4D_{5/2}$ state ($\ket{1}$), coupled by the 674 nm optical clock transition with a natural linewidth of 0.4 Hz.
    \textbf{c} Qubit State Preparation and Measurement (SPAM). Prior work was able to demonstrate qubit SPAM fidelity of 99.6$\%$ with trapped ion qubits using an integrated laser\cite{chauhan2026chip}, despite limited optical qubit coherence of only $T^*_2=$~60$\mu s$.
    \textbf{d}  Coherent Quantum Logic benchmarking. 
    In this new work, we implement multiple layers of phase noise mitigation which extend the optical qubit coherence time to $T^*_2=$~659 $\mu s$ and demonstrate single-qubit gates and two-qubit entanglement for the first time using an integrated photonic laser. 
    Randomized benchmarking yields a single-qubit fidelity of $\mathcal{F}_{1Q}=99.61\%\pm 0.03\%$ per Clifford gate and a two-qubit Mølmer-Sørensen interaction generates an entangled Bell state with a fidelity of $\mathcal{F}_{Bell}=92.35\% \pm 1.50\%$. 
	%
    }
	\label{fig:vision}
\end{figure*}

Here we demonstrate, for the first time, that an integrated photonic laser meets these requirements by driving single-qubit gates and two-qubit entanglement with an ultra-low noise Brillouin laser stabilized to a coil resonator, without the need for a table-top reference cavity. 
Randomized benchmarking\cite{knill2008randomized} yields an average single-qubit Clifford fidelity of $99.61\% \pm 0.03\%$, while a Mølmer-Sørensen\cite{sorensen2000entanglement} interaction generates a two-qubit entangled Bell state with fidelity of $92.35\% \pm 1.50\%$.  We additionally measure a bare optical-qubit coherence time of $T^*_2=$~659~$\mu s$ that extends to $T_2=$~1.75~ms with a spin echo.  These results demonstrate that the integrated photonic laser can be used for qubit state preparation, resolved sideband cooling, coherent single-qubit control, and two-qubit entanglement. These results are a critical step towards trapped ion quantum processors which monolithically integrate surface electrode ion traps with laser sources, light control, and beam delivery, creating a path for scalable, robust, and manufacturable trapped-ion based quantum computers.

\section{Results}

\subsection{Experiment}
The stabilized photonic laser used to drive the coherent optical qubit consists of a chip-scale Brillouin laser\cite{chauhanVisible780Nm2023,chauhan2021visible} that is Pound-Drever-Hall (PDH) locked to an integrated coil resonator\cite{chauhan2022integrated} (Figure \ref{fig:vision}a). This stabilized visible light Brillouin laser drives the \Sr optical qubit\cite{akerman2015universal} transition in a room temperature surface electrode trap (Figure \ref{fig:vision}b)\cite{chauhan2026chip}.
Both the laser and coil resonator are fabricated in an ultra-low loss \SiN platform\cite{blumenthalSiliconNitrideSilicon2018, chauhan2026chip,chauhanUltralowLossVisible2022} that is compatible with future monolithic integration with surface electrode\cite{seidelin2006microfabricated} ion-trap chips. In the present experiment, the Brillouin laser, coil resonator, and ion trap chip are separate chips connected by optical fiber and free-space optics which are susceptible to vibration- and polarization-induced noise. In this work we have mitigated these noise sources by stabilizing the beam path from the laser to the trapped ions using active fiber noise cancellation\cite{ma1994delivering}.
Future integration will reduce sensitivity to optical path perturbations, obviating the need for active fiber noise cancellation, simplifying experimental overhead while improving performance and robustness.
To suppress polarization-induced power fluctuations to the coil, we package the resonator with polarization-maintaining (PM) fiber aligned within a fiber-block array. This PM-fiber packaging stabilizes the input polarization, which in turn suppresses polarization-induced power fluctuations and therefore also the resulting thermally induced resonator frequency fluctuations.

\begin{figure*}
    \centering
    \includegraphics[width=\linewidth]{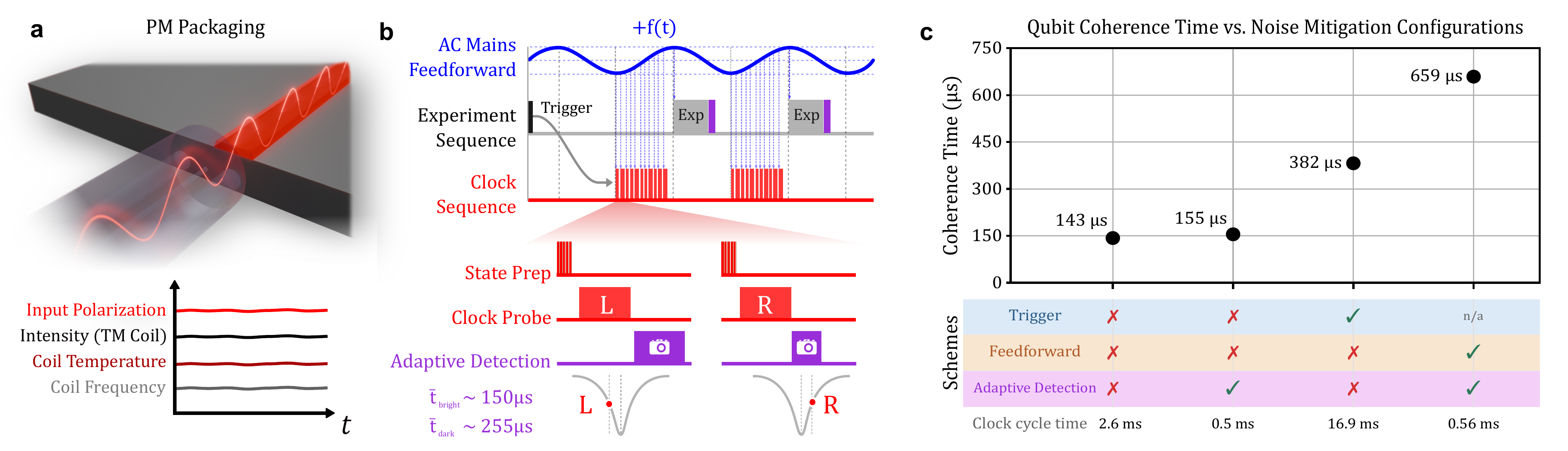}
    \caption{
    \textbf{Qubit coherence time improves through combined technical noise mitigation.
    }
    \textbf{a} Environmental and technical perturbations of the coil resonator are reduced with polarization maintaining fiber packaging. Stable input polarization reduces optical power fluctuations in the coil, reducing temperature fluctuations and thus frequency fluctuations of the coil.
    \textbf{b}~Timing sequence for stabilizing the integrated laser to the ion. The ion is interrogated on the left (L) and right (R) sides of the atomic resonance to generate an error signal which is fed back to the laser frequency. 
    Adaptive detection reduces the average bright and dark state detection times to $\sim 150~\mu s$ and $\sim 255~\mu s$ respectively, increasing the clock cycle rate and permitting multiple clock interrogations between experimental sequences. 
    Frequency feedforward $f(t)$ compensates for AC mains induced frequency shifts during all pulse sequences without trigger delays.
    \textbf{c} Qubit coherence time for different combinations of AC mains triggering, feedforward correction, and adaptive detection. 
    The simultaneous combination of adaptive detection and magnetic field frequency feedforward yields the largest improvement to qubit coherence times, enabling high-fidelity qubit gate operations. Trigger in this instance is not applicable (n/a) as the magnetic field feedforward is already referenced to the AC mains phase. All experiments used a PM packaged coil. Error bars are smaller than the markers. 
    }
    \label{fig:ClockTechniques}
\end{figure*}


We discipline the laser to the $\text{S}_{1/2} \leftrightarrow \text{D}_{5/2}$ transition in \Sr using an interleaved clock interrogation sequence\cite{chauhan2026chip} (see Fig. \ref{fig:ClockTechniques}b). This step compensates for thermal drift of the coil resonator between qubit experiments by interleaving clock probes, allowing the laser frequency delivered to the ion qubits to be continuously calibrated. 

The resulting lock performance determines whether the qubit coherence is limited by the intrinsic laser noise or by residual drift.
An independent optical-frequency-discriminator (OFD) measurement yields a reverse-integral linewidth of 580 Hz\cite{chauhan2026chip}, providing a target for laser-limited performance measured with the ion.
A linewidth near 500 Hz corresponds to a millisecond-scale Fourier limited interrogation time for the clock cycle. To optimize feedback from the clock sequence which disciplines the laser to the ion, the clock duty cycle should be maximized by minimizing the dead time from state preparation and measurement.
Additionally, at these millisecond timescales, optical fiber phase noise and AC mains induced magnetic field noise must also be suppressed, as the discipline feedback bandwidth is not sufficient to correct for fluctuations at these time scales.

\begin{figure*}
    \centering
    \includegraphics[width=\linewidth]{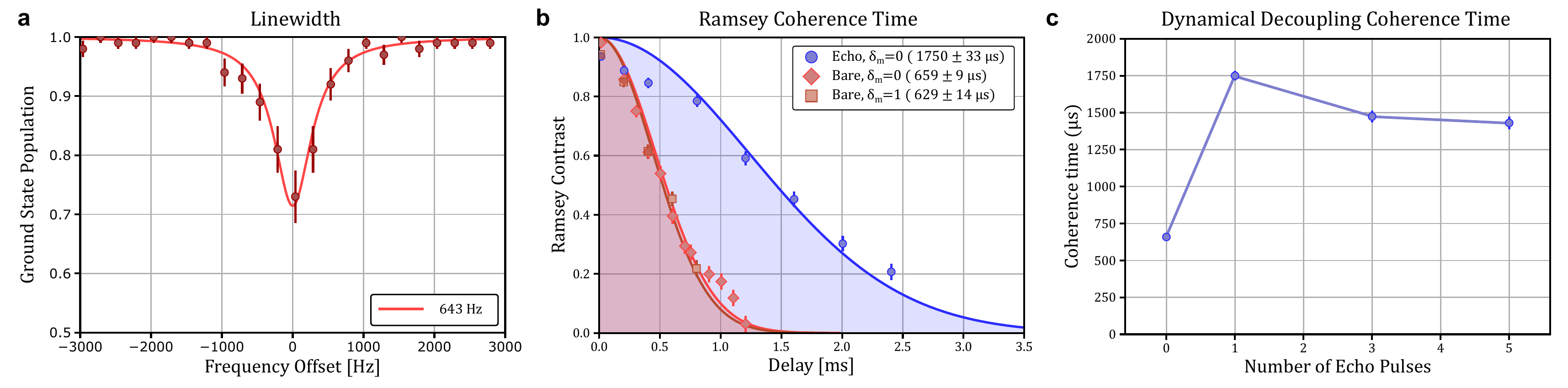}
    
    \caption{
    \textbf{Narrow linewidth qubit spectroscopy and Ramsey coherence time:}
    \textbf{a } Optical spectroscopy of the trapped ion qubit on the S to D transition driven with the integrated photonic laser gives a Lorentzian linewidth of $643\pm104\text{ Hz}$. Each data point consists of 100 shots with the full dataset collected over several minutes, indicating the long-term frequency stability of the integrated laser system. 
    %
    \textbf{b } Ramsey coherence measurements demonstrate more than an order-of-magnitude improvement to bare coherence time compared to our previous implementation\cite{chauhan2026chip}. Here we interrogate the 
    $\text{S}_{{1/2}, {+1/2}} \leftrightarrow \text{D}_{{5/2},{+1/2}}$ transition with our improved clock performance while feed-forwarding AC mains frequency shift and performing active fiber-noise cancellation. Fitting to a Gaussian decay envelope yields a bare $1/e$ coherence time of 659 $\mu\text{s}$, which extends to 1.75 ms with an echo.
    \textbf{c } Additional Uhrig Dynamical Decoupling (UDD) echo pulses do not extend the coherence time (solid line is a guide to the eye). 
    }
    \label{fig:LwAndCoherence}
\end{figure*}

\subsection{Adaptive Detection for Fast Clock Cycles}
High fidelity fluorescence detection typically requires a few milliseconds, introducing dead time comparable to the desired clock cycle probe time. Shortening detection therefore increases the clock duty cycle and therefore the bandwidth of the ion referenced frequency lock.
To speed up detection we use Bayesian inference to calculate the probabilities of a bright (S) state and dark (D) state every $20~ \mu s$ during fluorescence detection, halting detection as soon as a confidence threshold is reached\cite{Myerson_2008}.
The mean decision times are $150~ \mu s$ for bright states  and $255~ \mu s$ for dark states, compared to 1.44 ms for fixed duration threshold detection at matched detection fidelity.
This is a $9.6\times$ speedup for bright state detection and a $5.6\times$ speedup for dark state detection.


\subsection{Magnetic Field Software Feedforward}

The electrical current from the AC mains power generates periodic magnetic fields at 60 Hz, producing predictable Zeeman shifts in our trapped ions.
To characterize these shifts, we operate two independent clock feedback loops with interrogation pulses staggered by half a mains period.
We then slowly scan their common timing offset relative to the mains zero crossing to measure the differential frequency between each clock feedback loop, which precisely maps out the time-dependent Zeeman shift while rejecting common-mode laser drift.
A simultaneous sinusoidal fit to measurements on transitions with different magnetic susceptibilities gives an AC magnetic field amplitude of  $1.24 \pm 0.05$ mG (see Fig \ref{fig:SI_Bfield} in SI for details).
Using the calibrated frequency shift amplitude and the instantaneous phase of the mains cycle, we can then apply a real-time correction to the AOM frequency for every laser pulse to the ion. 
While higher-order harmonic corrections could also be made, they are smaller than our measured clock frequency deviations and therefore neglected. During preparation of this manuscript we became aware of related works\cite{debry2026real,tathed2026software} which correct for magnetic field noise via software and real-time calibrations.

Together, these techniques improve the precision with which the laser is disciplined to the ion, improving the qubit coherence time (Fig. \ref{fig:ClockTechniques}c). Combined with the improved stability provided by the PM-fiber packaging of the new coil resonator, these techniques extend the bare optical-qubit coherence time by more than an order of magnitude relative to our previous implementation\cite{chauhan2026chip}, enabling coherent qubit gate operations with the integrated photonic laser.

\subsection{Laser Noise and Qubit Coherence}
We first characterize the performance of the integrated laser using narrow linewidth optical spectroscopy and Ramsey interferometry. 
The 0.4 Hz natural linewidth of the $\text{S}\leftrightarrow\text{D}$ transition of \Sr allows the ion to serve as a sensitive probe of residual laser frequency noise, clock disciplining error, fiber path noise, and magnetic field noise. 
To characterize the effective laser linewidth experienced by the ion over several minutes, we perform spectroscopy on the narrow $\ket{S_{1/2}, m_J = +1/2} \rightarrow \ket{D_{5/2}, m_J = +1/2}$ clock transition with 1.3 ms long pulses. A Lorentzian fit gives a FWHM of $643 \pm 104 \text{ Hz}$ (see Fig. \ref{fig:LwAndCoherence}a). 
This linewidth is close to the 580 Hz reverse-integral linewidth independently measured with an OFD, indicating that the linewidth measured with the ion is approaching laser-limited performance.

We then characterize the relative laser-ion phase coherence using Ramsey interferometry\cite{langer2005long}.
The fringe contrast decays with a Gaussian envelope, from which we calculate the $1/e$ coherence time of the optical qubit controlled with our integrated photonic laser.
For the $S_{1/2,+1/2} \leftrightarrow D_{5/2,+1/2}$ transition (i.e. $\Delta_m=0)$, which has the lowest magnetic field sensitivity, we measure a decay time of $659 \pm 9$ $\mu s$ (see Fig. \ref{fig:LwAndCoherence}b), consistent with our measured linewidth.
For the $S_{1/2,+1/2} \rightarrow D_{5/2,-1/2}$ transition (i.e. $|\Delta_m|=1$), we measure a decay time of $629 \pm 14$ $\mu s$ (see Fig. \ref{fig:LwAndCoherence}b).
From these measurements and the known magnetic field susceptibilities of each transition\cite{barwood2015frequency} we can estimate the contribution of magnetic and non-magnetic noise to the decoherence. The coherence times of the two transitions being so similar despite over a factor of two different magnetic field susceptibilities suggests that non-magnetic dephasing dominates.


On the $S_{1/2,+1/2} \leftrightarrow D_{5/2,+1/2}$ transition, inserting an echo pulse driven by the integrated laser extends the qubit coherence time from 659~$\mu s$ to $1.750 \pm 0.033 \text{ ms}$, suggesting a substantial contribution from fixed frequency detuning offsets shot-to-shot. This would be consistent with imperfect disciplining of the laser being the dominant contribution to the bare qubit coherence time. We find that additional Uhrig dynamic decoupling (UDD)\cite{szwer2011keeping} echo pulses do not extend the coherence time further (see Fig. \ref{fig:LwAndCoherence}c).

\begin{figure*}
    \centering
    \includegraphics[width=1\linewidth]{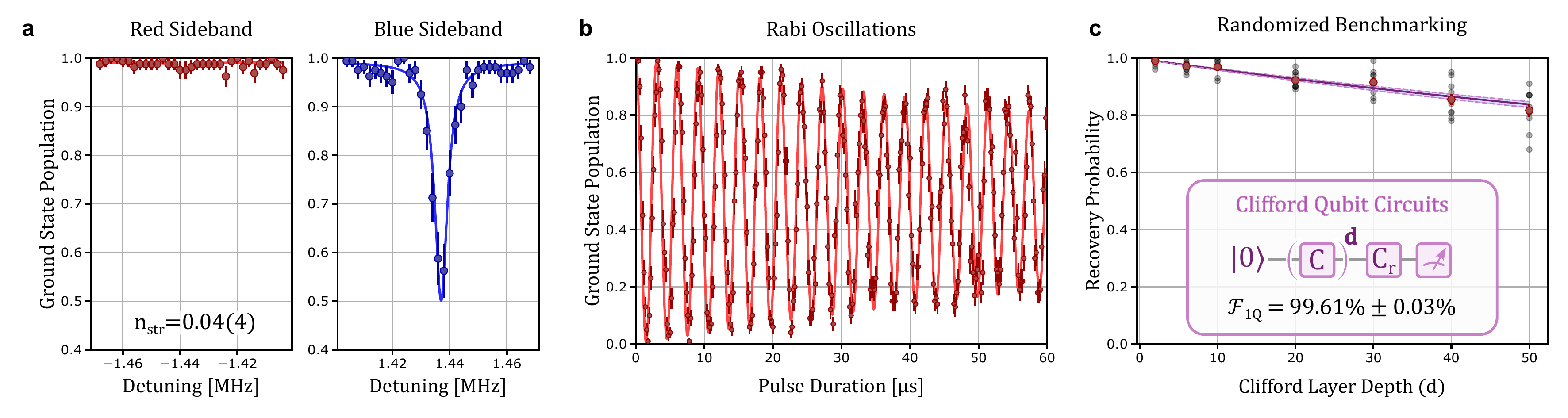}
    \caption{
    \textbf{Resolved Sideband Cooling and Single Qubit Gate Performance.}
    \textbf{a} Resolved sideband cooling showing $\bar{n}_{str}=0.04(4)$ of the out-of-phase motional mode (stretch mode) of two ions.
    %
    \textbf{b} Rabi oscillations on the $\ket{S_{1/2}, m_J = +1/2} \rightarrow \ket{D_{5/2}, m_J = +1/2}$ transition. The ion is cooled with Doppler laser cooling and polarization gradient cooling using a conventional laser, then the integrated laser is used to perform further resolved sideband cooling. We find a total residual thermal occupation of $\bar{n}\approx5.9(8)$ (including all motional modes) by fitting to a thermal model.
    \textbf{c} Randomized Benchmarking yields an average single qubit Clifford fidelity of $\mathcal{F}=99.61\pm0.03 \%$. For each depth, ten independently generated random Clifford sequences are measured with 100 shots each (grey markers), which are then averaged together (red markers). 
    Clifford gates are decomposed into native $\pi$ and $\frac{\pi}{2}$ rotations, with Z rotations implemented in software. The sequence depth $d$ denotes the number of Clifford gates.
    }
    \label{fig:QubitPerformance}
\end{figure*}

\subsection{Coherent Single Qubit Control}

We next evaluate the integrated laser as a control source for optical qubits\cite{akerman2015universal} through  resolved sideband cooling, coherent Rabi oscillations, and single-qubit randomized benchmarking\cite{knill2008randomized}.

After Doppler and polarization gradient cooling\cite{li2022robust}, which are performed using a conventional tabletop ECDL, we apply resolved sideband cooling\cite{monroe1995resolved} pulses to the two-ion axial stretch mode (i.e. out-of-phase mode) using our integrated laser. The sequence consists of steps with progressively longer red sideband pulses interleaved with optical pumping cycles, both of which are performed with the integrated laser. The optical pumping sequence also utilizes a separate tabletop DFB laser at 1033 nm which is used to repump the metastable D state to the excited P state, quenching any optical qubit population in the metastable D state to the ground S state in a few microseconds.
We then perform spectroscopy on the red and blue motional sidebands with the integrated laser and from the ratio of their amplitudes calculate a final thermal occupation of $\bar{n}_{str}=0.04(4)$ for our two-ion axial stretch mode (Fig. \ref{fig:QubitPerformance}a). By adding delays between the sideband cooling sequence and the final pulse probing the motional mode we can measure the heating rate. For the axial stretch mode we observe a heating rate of $\dot{\bar{n}}_{str}\sim$0.024~quanta/ms. 
We have also performed resolved sideband cooling to $\bar{n}\approx 1$ on the axial common mode (i.e. in-phase mode) and $\bar{n}\approx 0.3$ on radial common modes (i.e. in-phase mode), limited by heating rates of $\dot{\bar{n}}_{ax}\sim$5 quanta/ms (axial mode) and $\dot{\bar{n}}_{r_1}\sim$1 quanta/ms (lower radial mode), consistent with heating rates commonly observed in room temperature surface traps\cite{brownnutt2015ion}.



We first characterize the capability for the integrated laser to perform coherent qubit population transfer using Rabi oscillations and find that 90\% population inversion contrast is retained after twenty $\pi$ rotations (Fig.~\ref{fig:QubitPerformance}b).
At full power the $\pi$ pulse time is $T_{\pi}=1.5~\mu s$ and the oscillations damp out faster than the bare Ramsey coherence time, suggesting they are not laser limited. This damping is consistent with residual thermal motion, which creates shot-to-shot variations in the Rabi frequency, washing out the averaged contrast.
Fitting to a thermal model yields a total residual thermal occupation of $\bar{n}_{total}=5.9 \pm 0.8$.
Although individual modes can be cooled to lower motional states, cooling all modes simultaneously requires longer sequences as the red-sideband Rabi frequencies decrease near the motional ground state.
Because the thermal damping depends on the accumulated rotation angle, rather than the absolute Rabi frequency, lowering the optical power can extend the thermal damping timescale, allowing us to probe the optical-qubit coherence at longer times.
These slower Rabi oscillations exhibit an exponential decay time of 3.26~ms (see Fig \ref{fig:SI_long_time_scan}a).
Independent measurements of the delivered optical intensity show fluctuations of less than 1\%, suggesting intensity noise does not significantly contribute to the observed damping.

To further characterize the capability of the integrated laser for coherent qubit control, we measure the single-qubit Clifford fidelity by performing randomized gate benchmarking\cite{knill2008randomized}.
For each sequence depth $d$, we generate ten random Clifford sequences and perform 100 trials per sequence. After each sequence a final recovery Clifford ($C_r$) is applied which ideally returns the qubit to its initial state.
Fitting the average sequence fidelity for depths up to 50 Clifford layers to an exponential decay yields an average Clifford fidelity of $99.61 \pm 0.03 \%$ (Fig. \ref{fig:QubitPerformance}c) without SPAM corrections, see SI for details. 

\begin{figure*}
    \centering
    \includegraphics[width=\linewidth]{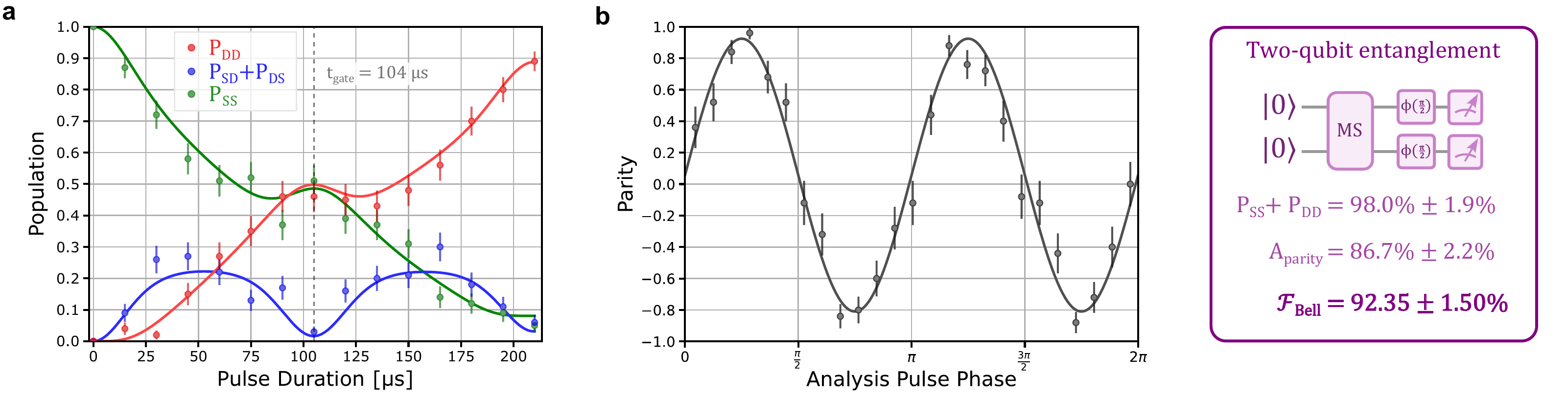}
    \caption{
    \textbf{Two-qubit Mølmer-Sørensen (MS) Gate.}
    \textbf{a} Time evolution of the MS interaction showing the even parity states $P_{SS}$ (green markers) and $P_{DD}$ (red markers) approach equal population at the gate time of $T_{gate}=104~\mu s$, while the combined odd parity population $(P_{SD}+P_{DS})$ (blue markers) reaches a minimum.
    Solid lines show a fit to a model with exponential coherence decay, yielding an effective decay time of 1.69~ms. 
    Each data point represents 100 shots, with error bars representing projection noise.
    \textbf{b} Parity oscillations $P=P_{SS}+P_{DD} - (P_{SD}+P_{DS})$, measured by scanning the phase of a final $\frac{\pi}{2}$ pulse applied after the MS gate time. 
    The fit amplitude of the parity contrast is $A=86.7\%\pm 2.2 \%$. Combining the parity contrast with the measured even-parity population of $98.0\%\pm 1.9 \%$ gives a Bell-state fidelity $\mathcal{F}_{Bell}= (P_{SS} + P_{DD} + A)/2 = 92.35\% \pm 1.50 \%$ with no SPAM corrections. Error bars represent projection noise over 100 shots.
    }
    \label{fig:MSgate}
\end{figure*}

\subsection{Two-Qubit Entangling Gate}
Two-qubit entangling gates impose additional requirements on laser phase noise near the ion motional frequencies.
To test the integrated laser in this more stringent regime, we perform two-qubit quantum logic gates using a Mølmer-Sørensen (MS) interaction\cite{sorensen2000entanglement, akerman2015universal}. Two laser tones, symmetrically detuned by 9.65 kHz from the red and blue sidebands of the two-ion axial stretch mode, are applied for 104 $\mu$s. A final phase-scanned $\pi/2$ pulse is then applied to both ions to measure the parity oscillations of the entangled state.
From these measurements, we extract a two-qubit Bell-state fidelity of $\mathcal{F}_{Bell}=92.35\% \pm 1.50\%$ (Fig. \ref{fig:MSgate}), without any SPAM corrections.
A fit (solid lines, see SI for details) yields an exponential decay time of 1.69~ms. Longer experimental time scans with continuous MS evolution over approximately six MS gates retain over $75\%$ contrast at 600~$\mu s$, fitting to an exponential decay time of 2.1~ms (see Fig. \ref{fig:SI_long_time_scan}b).

\section{Discussion}
We demonstrate the first coherent single-qubit and two-qubit quantum logic driven by a visible wavelength integrated laser stabilized to an integrated coil resonator, achieving a trapped ion optical qubit coherence time of 659 $\mu s$ (1.75 ms with spin echo), an average single-qubit Clifford fidelity of $99.61\%$, and two-qubit entangled state fidelity of $92.35\%$. These results demonstrate that integrated photonic lasers are not only capable of qubit state preparation and measurement, but also much more challenging applications requiring phase-coherent control such as single qubit gates and two qubit entanglement.
We show that it is possible to reach this level of qubit performance by addressing multiple sources of technical noise simultaneously.
These include mitigation of magnetic field noise by real-time frequency feedforward, active fiber-noise cancellation, polarization-maintaining fiber packaging, and faster clock feedback enabled by adaptive detection.
Future monolithic integration of the laser with the ion trap and modulation\cite{wangSiliconNitrideStressoptic2022} coupled with hybrid or heterogeneous integration of the laser semiconductor gain chip is expected to strongly suppress fiber-path and polarization induced noise due to the inherent stability of fixed on-chip waveguide routing and grating coupler delivery\cite{niffenegger2020integrated,mehta2020integrated}, improving qubit performance and robustness.
Integrating ion traps with laser systems could also enable scalable, distributed optical power generation and amplification. This could support parallel gate operations and maintain short gate times as trapped ion systems scale, which would not only reduce computational runtime but also preserve coherence during gates and therefore improve fidelity.

Improving the two-qubit Bell-state fidelity will require longer optical-qubit coherence.
The increase in coherence time from 659 $\mu s$ in bare Ramsey measurements to 1.75 ms with a spin echo indicates a substantial contribution from low-frequency, shot-to-shot detuning noise, consistent with residual errors in disciplining the laser to the ion.
In contrast, fitting the two-qubit MS dynamics with an exponentially decaying coherence envelope yields a coherence decay time of 1.69~ms, close to the independently measured spin-echo coherence time. This agreement is consistent with laser phase noise that persists under spin echo being a leading limitation for the MS interaction, rather than the slowly varying detuning noise originating from imperfect disciplining of the laser to the ion which dominates the bare Ramsey decay.
Extended measurements of continuous MS evolution over approximately six MS gates (see Fig. \ref{fig:SI_long_time_scan}b) are consistent with this observation, as they retain over $75\%$ contrast at 600~$\mu s$.
These longer scans are better fit by an exponential coherence decay envelope than a Gaussian decay and yield a coherence decay time of approximately 2.1~ms. Therefore, a fivefold increase in the coherence time could enable Bell-state fidelity of 99\%.

A direct way to improve the qubit coherence time would be to reduce the visible-wavelength propagation loss within the integrated coil resonator and use two point coupling approaches to optimize the coupling strength \cite{liu2024tunable}.  
This would narrow the resonator linewidth and optimize the extinction ratio to enable stronger suppression of laser frequency noise through the PDH lock, improving coherence.
Future integrated coil resonator designs employing low thermo-optic coefficient \cite{zhao2020low} and dual mode thermometry \cite{zhao2021integrated} approaches could additionally reduce thermal frequency drift, reducing the corrections required to discipline the laser to the ion and the associated residual detuning errors.


Improving the single qubit gate fidelity, however, will require improved ground state cooling. The decay of the single-qubit Rabi oscillations is consistent with residual thermal excitation of the ion, and is therefore not presently limited by the laser performance or qubit coherence.
The demonstrated single qubit gate performance is sufficient for shelving protocols used in optical-metastable-ground state (\textit{omg}) architectures\cite{allcock2021omg} which leverage the optical metastable state for multiple qubit encoding schemes. 
For instance, our demonstrated coherent qubit state transfer performance, both to and from the metastable D manifold, could support high fidelity heralded state detection protocols\cite{sotirova2024high}.

Integration of additional optical subsystems, particularly modulators, will be essential for realizing a fully integrated trapped ion quantum processor.
CMOS-compatible integrated optical modulators have been used to stabilize lasers to optical references\cite{wangSiliconNitrideStressoptic2022}. They have also  previously demonstrated amplitude modulation for trapped ion control\cite{hogle2023high}, but integrated frequency and phase modulation have not yet been demonstrated with trapped ions.
Conventional electro-optic platforms for phase modulation, such as thin-film lithium niobate (TFLN), may present challenges for full integration with ion-trap and \SiN photonic fabrication processes. 
Recently demonstrated visible-wavelength integrated lead zirconate titanate (PZT) phase modulators\cite{montifiore2026blue} with very low residual amplitude modulation (RAM), are a promising solution but have not yet been integrated with trapped ions.

In summary, we demonstrate that an integrated photonic laser can achieve the optical qubit coherence required for coherent single-qubit gates and two-qubit entanglement in a room temperature trapped ion system. 
This demonstrates that one of the most challenging optical subsystems required for trapped ion quantum computing, the narrow-linewidth stabilized laser, can be realized at chip-scale in a CMOS compatible \SiN platform with performance necessary for coherent qubit gates.  
This creates a path toward monolithic trapped ion quantum processors integrating lasers, modulators, optical routing, and beam delivery, establishing a foundation for fully integrated, scalable, and portable trapped ion quantum systems.

\section{Acknowledgments}

This work was supported in part by funding from the National Science Foundation under
Grant No. 2338369, Army Research Office (ARO) under award number W911NF2310179, and DARPA MTO award number FA9453-19-C-0030. The views, opinions and/or findings expressed are those of the author(s) and should not be interpreted as representing the official views or policies of the Department of Defense or the U.S. Government. The authors gratefully acknowledge help from Karl D. Nelson of Honeywell for chip fabrication. 

\section{Author Contributions}
 C.C., Z.W. and R.J.N. designed, fabricated, and packaged the ion trap surface electrode chip, performed the trapped ion qubit experiments, analyzed all data and wrote the manuscript. R.J.N. directed all experiments. J.D. implemented the fiber-noise cancellation. A. I., D. H., M.S., N.C., N.M. and K. L. designed, fabricated, tested and packaged the silicon nitride Brillouin laser and integrated coil resonator.  R.J.N. and D.J.B. conceived of the work and directed the research. All authors reviewed the manuscript.


\section{Competing interests}
D.J.B.’s work has been funded in the past by Infleqtion and he has consulted for Infleqtion and owns stock. All other authors declare no competing interests.

\section{Data availability}
The data that support the plots within this paper and other findings of this study are available from the corresponding author upon reasonable request.

\section{Code availability}
The code that supports the findings of this study is available from the corresponding authors upon reasonable request.

\section{Correspondence}
Correspondence and requests for materials should be addressed to R.J.N. (rniffenegger@umass.edu or \mbox{rniffene@ucr.edu}) and D.J.B. (danb@ucsb.edu).

\bibliography{ref}


\setcounter{figure}{0} 
\renewcommand{\thefigure}{S\arabic{figure}} 

~
\newpage
~

\begin{figure*}
    \centering
    \includegraphics[width=0.75\linewidth]{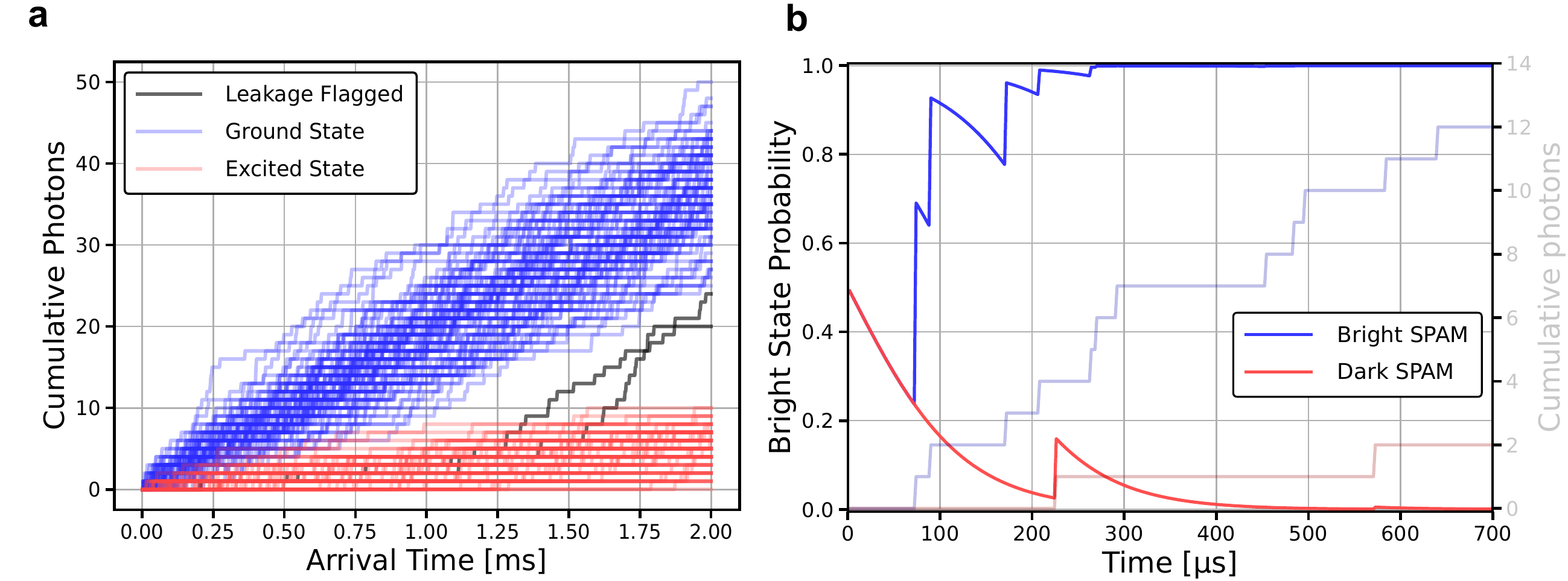}
    \caption{
    \textbf{Adaptive Bayesian Detection.}
    \textbf{a,} Cumulative photons collected from ion fluorescence vs arrival time, binned every 20 $\mu\text{s}$.
    After a conventional detection time of 2 ms it is possible to distinguish if the ion was prepared in the bright state (S) and dark state (D) with a total SPAM fidelity of $99.71\% \pm 0.08\%$.
    This time-resolved detection enables real-time Bayesian classification of the ion state, adaptively shortening detection time to speed up clock cycles. 
    \textbf{b,} Two sample trials from \textbf{a} plotted with the calculated probabilities vs. time. Photons are collected until a desired detection confidence is reached. Bright ions are detected $\times9.6$ faster (0.15 ms) and dark ions $\times5.6$ faster (0.255 ms) than conventional detection (1.44 ms), which thresholds only the final photon count.
    }
    \label{fig:SI_detection}
\end{figure*}

~
\newpage
\section{Supplemental Information}

\subsection{State preparation and measurement}

Time resolved Bayesian analysis also identifies spontaneous decay from the metastable D state during detection.
For example, in 2,017 trials prepared in the D state (dark state), there were 7 trials in which the ion was measured to be in the S state (bright), for a dark state SPAM fidelity of $\mathcal{F}_{SPAM,D}=99.65\pm0.1\%$. However, during these trials we identify 2 decay events (see Fig. \ref{fig:SI_detection}). Excluding these decay events over 4,000 total trials we measure a total SPAM fidelity of $\mathcal{F}_{SPAM}=99.71\%\pm0.08\%$.


\begin{figure}[b]
    \centering
    \includegraphics[width=0.8\linewidth]{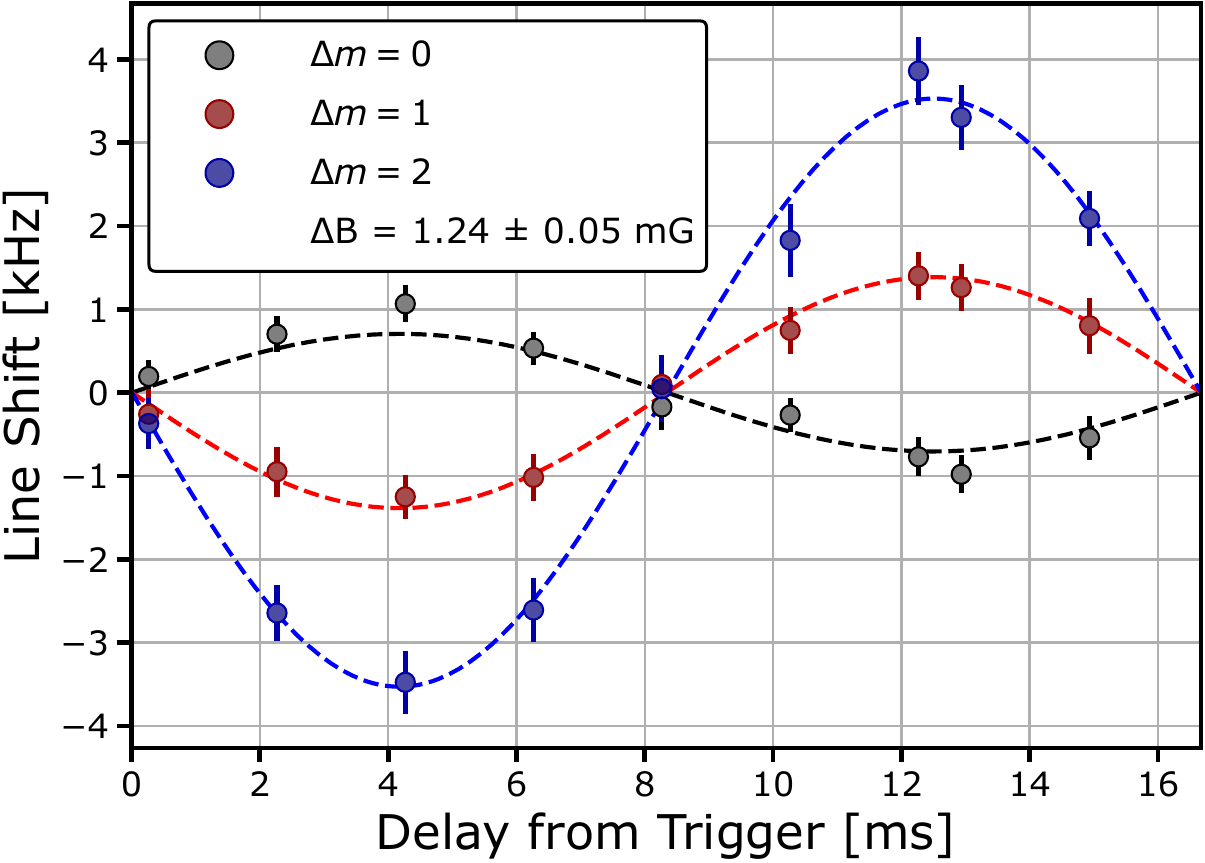}
    \caption{
    \textbf{AC line induced Zeeman Shifts.}
    Line shifts from magnetic fields generated by the AC mains for each spin transition. A global fit uses the known magnetic susceptibilities of the transitions and two free parameters: the magnetic field amplitude and phase. 
    Once calibrated we apply this shift as a `feedforward' correction to pulse frequencies at any time relative to the AC mains cycle.
    }
    \label{fig:SI_Bfield}
\end{figure}

\subsection{Magnetic Field Software Feedforward}

The AC mains magnetic-field noise fluctuations are superimposed on the 6.34 G static magnetic field used to resolve the Zeeman sub-levels.

To measure the magnetic field noise from the AC mains, we use 2 independent clock feedback loops, instanced in software separately, to enact phase-resolved differential sampling. The two clock routines occur with a relative delay of half a mains period. We collect 20 seconds of clock feedback data and then delay the global timing of both clock routines relative to the AC mains zero-crossing. By repeating this technique at various global delays, the difference in clock feedback maps out the relative magnetic field shift in time from AC mains.

A simultaneous sinusoidal fit to measurements on transitions with different magnetic susceptibilities gives an AC magnetic field amplitude of  $1.24 \pm 0.05$ mG, corresponding to peak-to-peak line shifts of -1.41 kHz ($\text{S}_{{1/2},+{1/2}}\leftrightarrow\text{D}_{{5/2},+{1/2}}$), 2.77 kHz ($\text{S}_{{1/2},+{1/2}}\leftrightarrow\text{D}_{{5/2},+{3/2}}$), 7.06 kHz ($\text{S}_{{1/2},+{1/2}}\leftrightarrow\text{D}_{{5/2},+{5/2}}$), see \mbox{Figure \ref{fig:SI_Bfield}}.


\subsection{Randomized Gate Benchmarking Analysis}


For each depth in our randomized gate benchmarking data, we average the probability of successful recovery to the original qubit state (S state) over 1000 trials (100 shots of 10 unique circuits). We plot this sequence success fidelity $F$, versus the circuit depth \textit{d}, and extract the error-per-Clifford ($\epsilon$) from a fit to $F(d) = \alpha (1-2\epsilon)^{d} + 0.5$. The reported Clifford fidelity is given by $1-\epsilon$, with the $\pm1\sigma$ bounds obtained from the $\chi^2$ evaluation.


\subsection{Mølmer-Sørensen model}

\begin{figure*}
    \centering
    \includegraphics[width=0.9\linewidth]{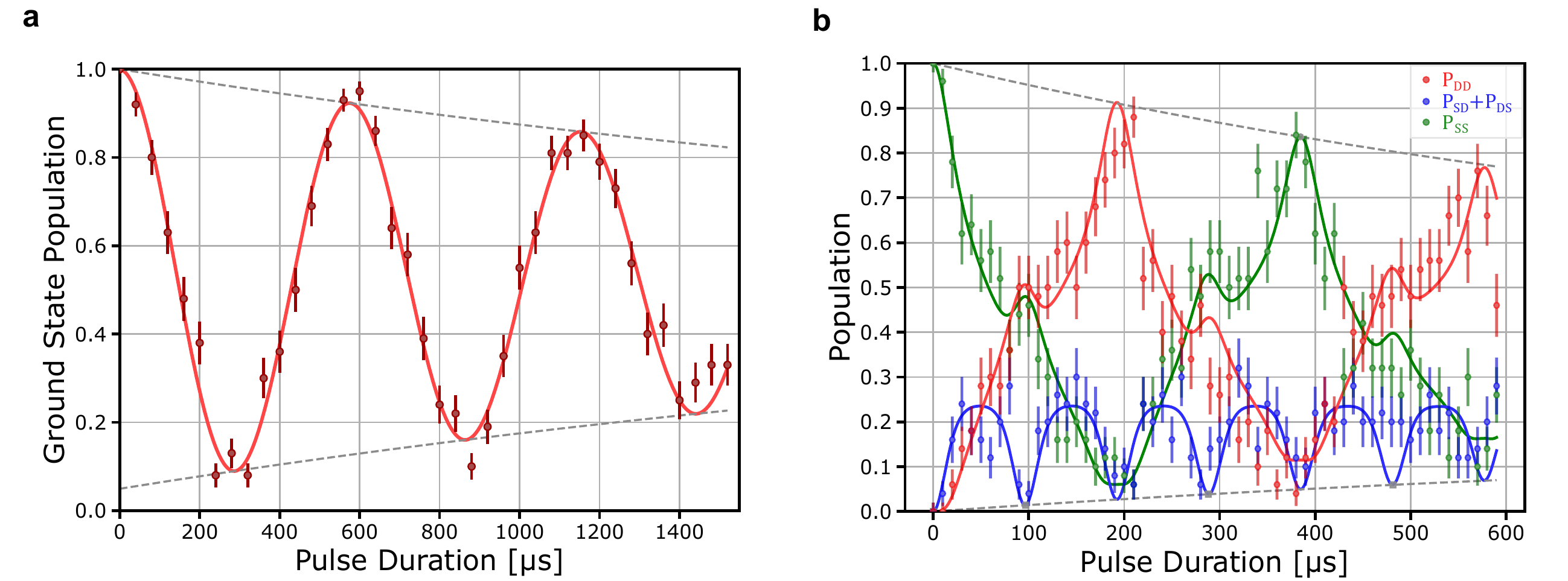}
    \caption{
    \textbf{Long qubit time scans.}
    \textbf{a} Low power optical pulses allow us to measure the coherence decay of Rabi oscillations at longer time scales, which reduces the sensitivity to motional and intensity related dephasing effects. A fit to an exponential decay envelope (dashed lines) yields a coherence decay time of $3.26$~ms.
    \textbf{b} Longer time scans of an MS gate interaction also show an exponential decay envelope (dashed lines) which fits to a $1/e$ coherence decay time of 2.08~ms, and symmetric detuning error of $\delta_{sym}= 156$~Hz.
    %
    }
    \label{fig:SI_long_time_scan}
\end{figure*}

We use QuTiP\cite{qutip5} to model the MS gate interaction.
The phase noise is modeled with a Lindblad collapse operator for collective dephasing to extract the exponential coherence decay times (Fig. \ref{fig:SI_long_time_scan}). The model assumes that the heating rate of the out-of-phase axial ('stretch') mode is zero, as our heating rate measurement of 0.024 q/ms, suggests that during a 100 $\mu s$ gate there would be less than 0.01 quanta added. Our fits allow for nonzero symmetric and asymmetric detunings to fully capture population dynamics.


~

\end{document}